\pdfoutput=1
\documentclass[reprint,prl,aps,twocolumn,preprintnumbers,nofootinbib,superscriptaddress,showpacs,floatfix]{revtex4}
\usepackage{amsmath,amssymb,amsbsy}
\usepackage{graphicx}
\usepackage{epsfig}
\usepackage{color}

\newcommand{\Delstar}{\ensuremath{\Delta^{\raise0.18ex\hbox{${\scriptstyle *}$}}}}
\def\gtwid{{\,\raise.35ex\hbox{$>$\kern-.75em\lower1ex\hbox{$\sim$}}\,}}
\def\ltwid{{\,\raise.35ex\hbox{$<$\kern-.75em\lower1ex\hbox{$\sim$}}\,}}
\def\leftvec{{\raise1.5ex\hbox{$\leftarrow$}\kern-1.00em}}
\def\rightvec{{\raise1.5ex\hbox{$\rightarrow$}\kern-1.00em}}
\def\half{{\scriptstyle \raise.2ex\hbox{${1\over2}$}}}
\def\threehalves{{\scriptstyle \raise.15ex\hbox{${3\over2}$}}}
\def\third{{\scriptstyle \raise.15ex\hbox{${1\over3}$}}}
\def\third{{\scriptstyle \raise.15ex\hbox{${1\over3}$}}}
\def\twothirds{{\scriptstyle \raise.15ex\hbox{${2\over3}$}}}
\def\fourth{{\scriptstyle \raise.15ex\hbox{${1\over4}$}}}

\newcommand*{\bea}{\begin{eqnarray}}
\newcommand*{\eea}{\end{eqnarray}}
\newcommand*{\be}{\begin{equation}}
\newcommand*{\ee}{\end{equation}}

\newcommand*{\CPT}{\raise0.4ex\hbox{$\chi$}PT}
\newcommand*{\chpt}{\raise0.4ex\hbox{$\chi$}PT}
\newcommand*{\schpt}{S\raise0.4ex\hbox{$\chi$}PT}

\def\eqref#1{{(\ref{#1})}}

\def\bar{\overline}
\def\hat{\widehat}

\def\bea{\begin{eqnarray}}
\def\eea{\end{eqnarray}}

\def\beq{\begin{equation}}
\def\eeq{\end{equation}}

\def\spose#1{\hbox to 0pt{#1\hss}}
\def\ltapprox{\mathrel{\spose{\lower 3pt\hbox{$\mathchar"218$}}
 \raise 2.0pt\hbox{$\mathchar"13C$}}}
\def\gtapprox{\mathrel{\spose{\lower 3pt\hbox{$\mathchar"218$}}
 \raise 2.0pt\hbox{$\mathchar"13E$}}}
\def\inapprox{\mathrel{\spose{\lower 3pt\hbox{$\mathchar"218$}}
 \raise 2.0pt\hbox{$\mathchar"232$}}}

\begin{document}

%\preprint{IUHET-525}

%\vphantom{}

\title{Unitarity Triangle {\em Without} Semileptonic Decays}

\author{E.\ Lunghi}
%\email[]{elunghi@indiana.edu}
\affiliation{Physics Department, Indiana University, Bloomington, IN 47405}

\author{Amarjit Soni}
%\email[]{soni@bnl.gov}
\affiliation{Physics Department, Brookhaven National Laboratory, Upton, NY 11973, USA }

%=================================================
%The abstract
%=================================================
\begin{abstract}
Use of semi-leptonic decays has become standard in constraining the Unitarity Triangle. Bearing in mind that precise calculations of these are very challenging, we propose an entirely new approach. In particular the $|V_{cb}|$ + $\varepsilon_K$ constraint, which depends extremely sensitively on $|V_{cb}|$ in the traditional method, is replaced by the interplay between $\varepsilon_K$, ${\rm BR}(B\to\tau\nu)$ and $\Delta M_{B_s}$. It is found that even in this method tensions with the Standard Model persist at the $\sim1.8 \sigma$ level. Furthermore, improvements on the $B\to\tau\nu$ branching ratio and on the lattice determination of $f_{B_s} \hat B_s^{1/2}$ can increase the effectiveness of this method significantly.
\end{abstract}
\maketitle
%
%\paragraph{\bf Introduction.}
%
The next big step in our understanding of particle physics will be the uncovering of the electro-weak symmetry breaking (EWSB) mechanism. The present and upcoming collider experiments (Fermilab and LHC) will be able to test the Standard Model (SM) Higgs mechanism. New physics is widely expected at around the TeV scale if the Higgs mass is not to receive large radiative corrections and require severe  fine-tuning. A stringent constraint on the SM mechanism of EWSB is the tight structure of flavor changing (FC) interactions: tree--level FC neutral currents are forbidden and charged currents are controlled by the Cabibbo--Kobayashi--Maskawa (CKM)~\cite{ckm} mixing matrix
\bea
V & = & 
\begin{pmatrix}
1-\frac{\lambda^2}{2} & \lambda & A \lambda^3 (\rho- i \eta)\cr
-\lambda & 1-\frac{\lambda^2}{2} & A \lambda^2 \cr
A \lambda^3 (1-\rho-i \eta) & - A \lambda^2 & 1 \cr
\end{pmatrix} . \nonumber
\eea
Within the SM, the CKM matrix is the only source of FC interactions and of CP violation. There is no reason, in general, to expect that new physics (needed to stabilize the Higgs mass) at the TeV scale will be in the basis wherein the quark mass matrix is diagonal. This reasoning gives rise to another fundamental problem in particle physics, namely the flavor puzzle i.e. unless the scale of new physics is larger than $10^3$ TeV it causes large FCNC especially for the $K-\bar K$ system. Thus flavor physics provides constraints on models of new physics upto scales that are much much larger than what is accessible to direct searches at colliders such as  the Tevatron or the LHC. Flavor physics is therefore expected to continue to provide crucial information for the interpretation of any physics that LHC may find.

In this decade significant progress has been made in our understanding of flavor physics, thanks in large part to the spectacular performance of the two asymmetric B-factories. For the first time, we learned that the CKM-paradigm of CP violation is able to simultaneously account for the observed CP-violation in the K and B-systems up to an accuracy of O(20\%). Impressive as this is, it is still important to understand that this leaves a lot of room for new physics. Indeed, as more data from B-factories became available and also as the accuracy in some key theoretical calculations was attained, several 2-3 $\sigma$ hints of new physics have emerged. While this clearly does not represent an unambiguous signal for new physics, it does mean that efforts need to continue  both on the experiment and on the theory front to seek greater clarity with regard to these anomalies. 

In this context use of semileptonic decays in all traditional analysis of the Unitarity Triangle (UT) to date is a concern. The inclusive  $b \to u$  transitions are not governed by any symmetry and as a result are a special challenge for continuum methods. Exclusive decays are in principle amenable to the lattice and steady, but unfortunately rather slow, progress is being made. The fact that for both $b \to c$ and for $b \to u$ inclusive and exclusive methods disagree by $\approx 2 \sigma$ casts a shadow of doubt on the results of the UT analysis. This is especially aggravated by the fact that use of the input from $\epsilon_K$, representing the indirect CP violation from the $K_L \to \pi \pi$, into the UT is exceedingly sensitive to $V_{cb}$, scaling as the fourth power. These observations motivate us to seek alternate approaches, which we will provide herein, though we want to emphasize that we are not suggesting that efforts to improve our understanding of semi-leptonic decays be discontinued but rather that alternate methods of analysis for the UT could be very valuable and should also be developed. 
 
Recent improvements especially in the lattice calculation of $B_K$~\cite{Gamiz:2006sq,Antonio:2007pb,Aubin:2009jh,RBC-UKQCD09}, led to the appearance of a $\sim 2 \sigma$ tension that can be interpreted as new physics in $B_d$ and/or in $K$ mixing~\cite{Lunghi:2007ak, Lunghi:2008aa, Buras:2008nn, Buras:2009pj, Lunghi:2009sm}. The interplay of $\varepsilon_K$ with $\Delta M_{B_s}/\Delta M_{B_d}$ and $S_{\psi K}$ (time--dependent CP asymmetry in $B\to J/\psi K_s$) is at the heart of the tension. The inclusion of $|V_{ub}|$ from semileptonic $b\to u\ell \nu$ decays, tends to favor a scenario with new physics in kaon mixing~\cite{Laiho:2009eu}. An important difficulty with these analyses is the long standing discrepancy between the extraction of $|V_{cb}|$ and $|V_{ub}|$ from exclusive and inclusive semileptonic decays alluded to above. From the inspection of Table~\ref{tab:utinputs}, one sees that inclusive and exclusive determinations differ at the $~2\sigma$ level. While Ref.~\cite{Lunghi:2008aa} demonstrated that $|V_{ub}|$ can be dropped from the fit without affecting the observed tension, it is usually believed that $|V_{cb}| \simeq A \lambda^2$ from semileptonic decays is essential in order to use $\varepsilon_K$ (because of its $A^4$ dependence).

Bearing all this in mind, in this letter we propose a new approach to the  UT analysis, wherein no use of semi-leptonic decays is made. In particular, we show that the traditional use of $\varepsilon_K$ + $|V_{cb}|$ combination can be effectively replaced by the interplay between $\varepsilon_K$, $\Delta M_{B_s}$ and ${\rm BR} (B\to \tau \nu)$. We find that even after removing all input to UT analysis  from semileptonic decays, the 2$\sigma$ tension in the UT fit with the SM survives. More importantly, every experimental and theoretical input to this analysis is now theoretically clean and  under very good control. The latter point is quite important, because many of the hints for new physics that come from precision studies tend to have some problems. For instance, the muon anomalous magnetic moment ($(g-2)_\mu$) tension relies on the non-perturbative estimation of light--by--light scattering contributions, the use of $e^+ e^- \to {\rm hadrons}$ to calculate the hadronic vacuum polarization tensor of the photon and on isospin corrections to $\tau$ decays data. Another example is the $(2-3)\sigma$ tensions in the direct CP asymmetries in $B\to K\pi$ as well as in time--dependent CP asymmetries  $B\to \psi K$ versus the penguin modes such as $B \to \phi K $ or $\eta^\prime K$ etc. These rely on model dependent QCD treatment of power corrections and non-factorizable effects which are very difficult to ascertain reliably. A very important exception is the $2.2\sigma$ evidence for a CP violating phase in $B_s$ mixing, whose non-zero value would be an extremely clean evidence for new physics~\cite{Lenz:2006hd, Aaltonen:2007he, :2008fj, Bona:2008jn}. 
%
%%%UPDATE%%%
%
\begin{table}[t]
\begin{center} 
\begin{tabular}{ll}
\toprule
$\left| V_{cb} \right|_{\rm excl} = (38.6 \pm 1.2) 10^{-3}$& 
$\eta_1 = 1.51 \pm 0.24$ \cr
$\left| V_{cb} \right|_{\rm incl} = (41.31 \pm 0.76) 10^{-3}$&  
$\eta_2 = 0.5765 \pm 0.0065$\cr
$\left| V_{cb} \right|_{\rm incl + excl} = (40.3 \pm 1.0) 10^{-3}$  & 
$\eta_3 = 0.47 \pm 0.04$ \cr
$\left| V_{ub} \right|_{\rm excl} = (34.2 \pm 3.7) 10^{-4}$  &
$\eta_B = 0.551 \pm 0.007$\cr
$\left| V_{ub} \right|_{\rm incl} =  (40.1 \pm 2.7 \pm 4.0) 10^{-4} $    & 
$\xi = 1.23 \pm 0.04$\cr
$\left| V_{ub} \right|_{\rm incl + excl} = (36.4 \pm 3.0) 10^{-4}$  & 
$\lambda = 0.2255  \pm 0.0007$ \cr
$\Delta m_{B_d} = (0.507 \pm 0.005)\; {\rm ps}^{-1}$ & 
$\alpha = (89.5 \pm 4.3)^{\rm o}$\cr
$\Delta m_{B_s} = (17.77 \pm 0.12 )\;  {\rm ps}^{-1}$ & 
$S_{\psi K_S} = 0.672 \pm 0.024$\cr
$\varepsilon_K = (2.229 \pm 0.012 ) \times 10^{-3}$&
$\gamma = (78 \pm 12)^{\rm o}$ \cr
$m_c(m_c) = (1.268 \pm 0.009 ) \; {\rm GeV}$ &
$\hat B_K = 0.725 \pm 0.027$ \cr
$m_{t, pole} = (172.4 \pm 1.2) \; {\rm GeV}$ & 
$f_B = (192.8 \pm 9.9) \; {\rm MeV}$ \cr
$f_{B_s} \sqrt{\hat B_s}  =  (275 \pm 19) \; {\rm MeV}$ &
$f_K = (155.8 \pm 1.7) \; {\rm MeV}$ \cr
${\cal B}_{B\to \tau\nu} = (1.43\pm 0.37) 10^{-4}$\cite{hfag} &
$\kappa_\varepsilon = 0.92 \pm 0.01$  \cr
\botrule
\end{tabular}
\caption{Inputs used in the fit. References to the original experimental and theoretical papers and the description of the averaging procedure can be found in Ref.~\cite{Laiho:2009eu}. Statistical and systematic errors are combined in quadrature. We adopt the averages of Ref.~\cite{Laiho:2009eu} for all quantities with the exception of $|V_{ub}|$, $\xi$ and $f_{B_s} \sqrt{\hat B_s}$ (see text). \label{tab:utinputs}}
\end{center}
\end{table}
\paragraph{\bf Present status of the UT fit.}
\label{sec:sm}
The standard technique to extract the parameters $A$, $\rho$ and $\eta$ relies on a simultaneous fit of the  following observables: the mixing induced CP violation in the kaon system ($\varepsilon_K$), the mass differences in the $B_d$ and $B_s$ systems ($\Delta M_{B_{d,s}}$), the CP asymmetries in $B\to J/\psi K_s$ ($S_{\psi K}$), $B\to (\pi\pi, \rho\rho, \rho\pi)$ ($\alpha$) and $B\to D^{(*)} K^{(*)}$ ($\gamma$) modes, the rates of semileptonic $b\to (c,u) \ell \nu$ decays ($|V_{cb}|$ and $|V_{ub}|$)~\cite{footnote1} and the $B\to \tau \nu$ branching ratio. A complete description of these observables and of the possible statistical techniques that could be employed can be found in Refs.~\cite{Charles:2004jd,Ciuchini:2000de}. We follow the approach of Refs.~\cite{Lunghi:2008aa,Lunghi:2009sm} and utilize the averages calculated in Ref.~\cite{Laiho:2009eu} with some exceptions: we include inclusive $|V_{ub}|$ albeit with an additional $10\%$ model uncertainties~\cite{Bona:2009tn}; we take a simple (not weighted) average of the determinations of $\xi$ from Fermilab/MILC~\cite{Evans:2008zz} and HPQCD~\cite{Gamiz:2009ku}. Also we take the central value of $f_{B_s} \sqrt{B_s}$ from Ref.~\cite{Laiho:2009eu} but adopt the uncertainty quoted in Ref.~\cite{Gamiz:2009ku}. We adopt this conservative stance to show that the impact of the approach we champion in this letter remains largely unaffected even if the lattice errors are not as small as currently claimed in the literature.

We summarize the inputs we use in Table~\ref{tab:utinputs}. Below we first present explicitly only those formulas that are relevant to the traditional analysis which uses semi-leptonic decays:
\begin{figure}[t]
\begin{center}
\includegraphics[width=0.85 \linewidth]{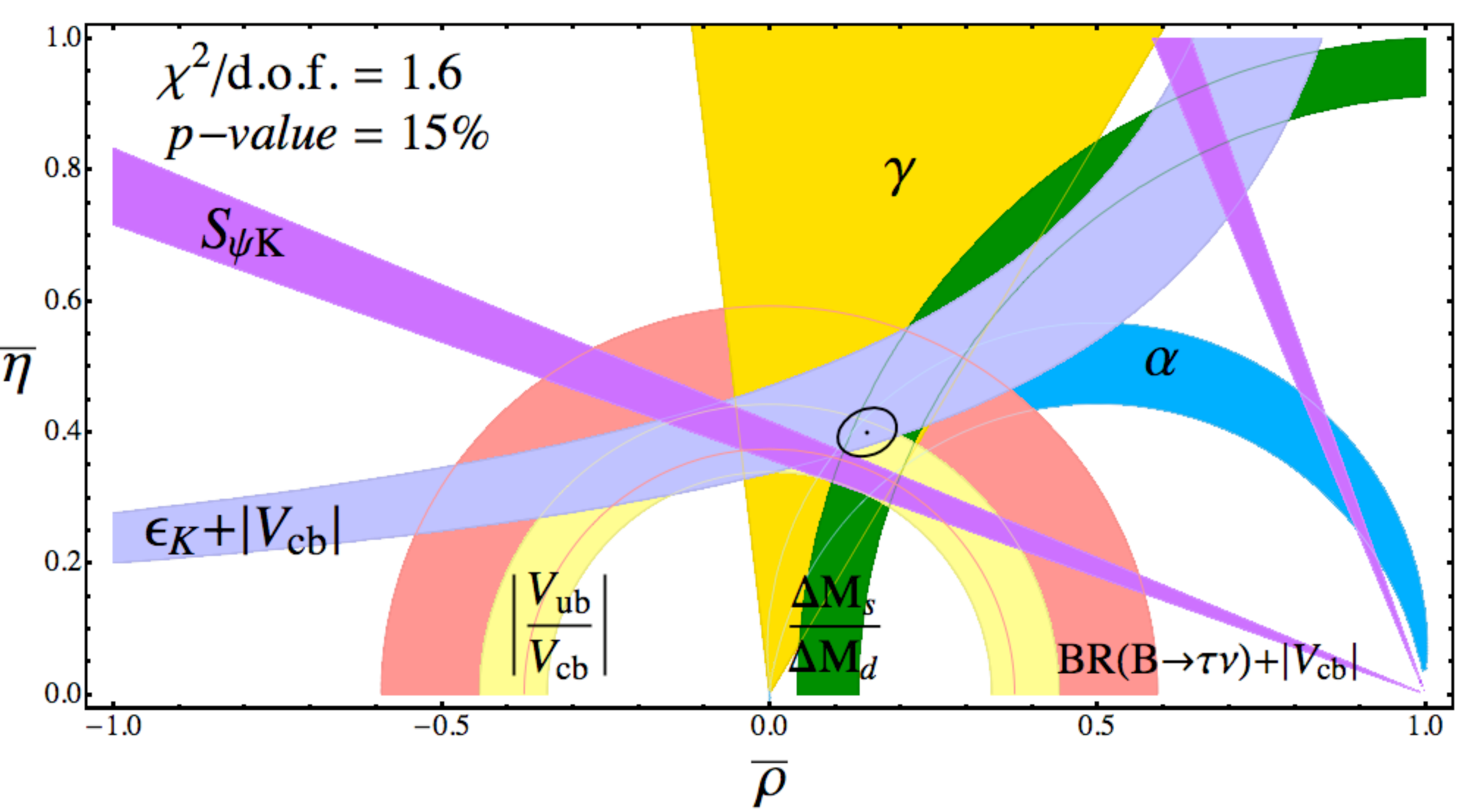}
\caption{Standard unitarity triangle fit. The contour is obtained using $V_{cb}$, $V_{ub}$, $\varepsilon_K$, $B\to\tau\nu$, $\gamma$, $\Delta M_{B_s}$ and $\Delta M_{B_d}$.\label{fig:utfitstandard}}
\end{center}
\end{figure}
{
\allowdisplaybreaks
\bea
\Delta M_{B_s} & =&   \chi_s\; f_{B_s}^2  \hat B_{B_s} A^2 \lambda^4   \label{dmbs} \\
\Delta M_{B_d} &=&   \chi_d\; f_{B_d}^2  \hat B_{B_d} A^2 \lambda^6 ( \eta^2 + (-1 + \rho)^2 ) \label{dmbd}  \\
\frac{\Delta M_{B_s}}{\Delta M_{B_d}} & =&  \frac{m_{B_s}}{m_{B_d}} \frac{ \xi^2 \lambda^{-2}}{ \eta^2 + (-1 + \rho)^2 } \label{xsd} \\
|\varepsilon_K |  &=& 2 \chi_\varepsilon    \hat B_K  \kappa_\varepsilon \; \eta \lambda^6 \Big(
 A^4 \lambda^4 (\rho-1) \eta_2 S_0 (x_t) \nonumber \\
&& +A^2 \big( \eta_3 S_0 (x_c,x_t) -\eta_1 S_0 (x_c) \big)  \Big)  \label{ek}\\
&&  \hskip -2 cm  {\rm BR} (B\to \tau \nu) = \chi_\tau  f_B^2 \left| V_{ub} \right|^2 \simeq \chi_\tau  f_B^2 
A^2 \lambda^6 (\rho^2 + \eta^2)
\label{btn}
\eea
}
where we expanded in $\lambda$ and defined 
\bea
\chi_q &=& G_F^2 m_W^2 m_{B_q} \eta_B S_0 ( x_t)/(6  \pi^2) \; , \\
\chi_\varepsilon &=& (G_F^2 m_W^2 f_K^2 m_K)/(12 \sqrt{2} \pi^2 \Delta m_K^{\rm exp}) \; , \\
\chi_\tau &=&  G_F^2 m_\tau^2 m_{B^+}/(8\pi \Gamma_{B^+}) ( 1- m_\tau^2 /m_{B^+}^2)^2.
\eea
The $68\%$ C.L. allowed regions in the $(\rho,\eta)$ plane are shown in Fig.~\ref{fig:utfitstandard}, where we show explicitly that the $\varepsilon_K$, $B\to\tau\nu$ (pink) and $|V_{ub}|$ (yellow) constraints require $|V_{cb}|$ in order to be drawn independently. In particular we obtain:
%%%UPDATE%%%
%
\bea
|V_{ub}|_{\rm fit} & = & ( 3.61 \pm 0.13 ) \times 10^{-3} \, , \label{vubfit}\\
{\rm BR} (B\to\tau\nu)_{\rm fit} & = & ( 0.87 \pm 0.11 ) \times 10^{-4} \, , \label{brbtnfit}\\
\left[ \sin 2\beta \right]_{\rm fit} & = & 0.766 \pm 0.036  \, , \label{sin2betafit}
\eea
where each quantity is extracted by removing the corresponding direct determination form the fit (for $V_{ub}$ this means removing information from semileptonic $b\to u \ell \nu$ decays~\cite{footnote1}).
%%%UPDATE p-values (3)%%%
Note that $|V_{ub}|_{\rm fit}$ is quite close to the determination from exclusive decays and that ${\rm BR} (B\to\tau\nu)_{\rm fit}$ is {\em smaller} than the corresponding world average ($(1.43 \pm 0.37) \times 10^{-4}$~\cite{hfag}). The result in Eq.~(\ref{brbtnfit}) is driven by the updated value for the $B$ decay constant. The relatively low p--value~\cite{footnote2} ($p = 15\%$) of the overall fit has been interpreted in terms of new physics in either $K$ or in $B_d$ mixing~\cite{Lunghi:2007ak,Lunghi:2008aa,Buras:2008nn,Buras:2009pj,Lunghi:2009sm,Laiho:2009eu}.

Adopting the model independent parametrizations
\bea
|\varepsilon_K^{\rm NP}| &=& C_\varepsilon \; |\varepsilon_K^{\rm SM}| \; , \\
M_{12}^{d,{\rm NP}} &=& e^{i \theta_d} \; M_{12}^{d,{\rm SM}} \;, \\
{\rm BR} (B\to \tau\nu)^{\rm NP} &=& r_H \;  {\rm BR} (B\to\tau\nu)^{\rm SM}  \; ,
\eea
where $M_{12}^d$ is the matrix element of the effective Hamiltonian between $B_d$ and $\bar B_d$ meson states and in the SM we have ($C_\varepsilon, \; r_H )= 1$ and $\theta_d = 0$. We obtain
%%%UPDATE%%%
%
\bea
C_\varepsilon & = & 1.28 \pm 0.14 \hskip 0.45 cm \Rightarrow \; (2.0\sigma, p = 58\%) \; , \label{cepsilonfull}\\
\theta_d & = & -(4.0 \pm 1.8)^{\rm o}\; \Rightarrow   \; (2.2\sigma, p= 61\%) \; , \label{thetadfull}\\
r_H & = & 1.7 \pm 0.5 \hskip 0.82cm \Rightarrow \; (1.4\sigma, p = 29\%) \; .
\eea
%
%%%UPDATE p-value%%%
The above results are mutually exclusive in the sense that they are obtained by allowing either new physics in $K$, or in $B_d$ mixing or in $B_u \to \tau \nu$ and point to a $\sim 2 \sigma$ hint for new physics. The quoted p--values have to be compared with SM result ($ p_{\rm SM} = 15\%$). The lower p--value for new physics in $B\to \tau\nu$ indicates that the tension in the fit is only partially lifted by new contributions to $B\to\tau\nu$.
\begin{figure}[t]
\begin{center}
\includegraphics[width=0.85 \linewidth]{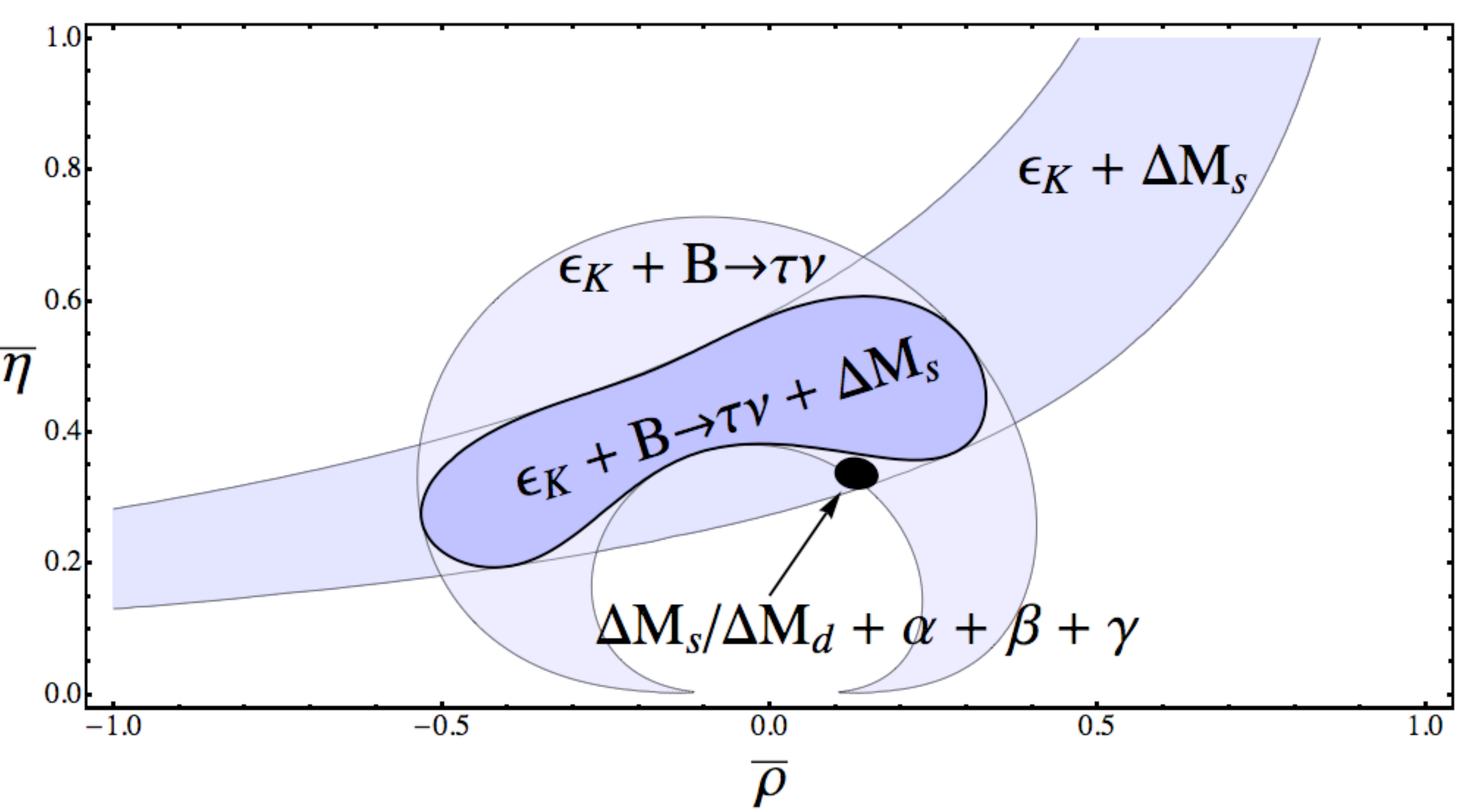}
\includegraphics[width=0.85 \linewidth]{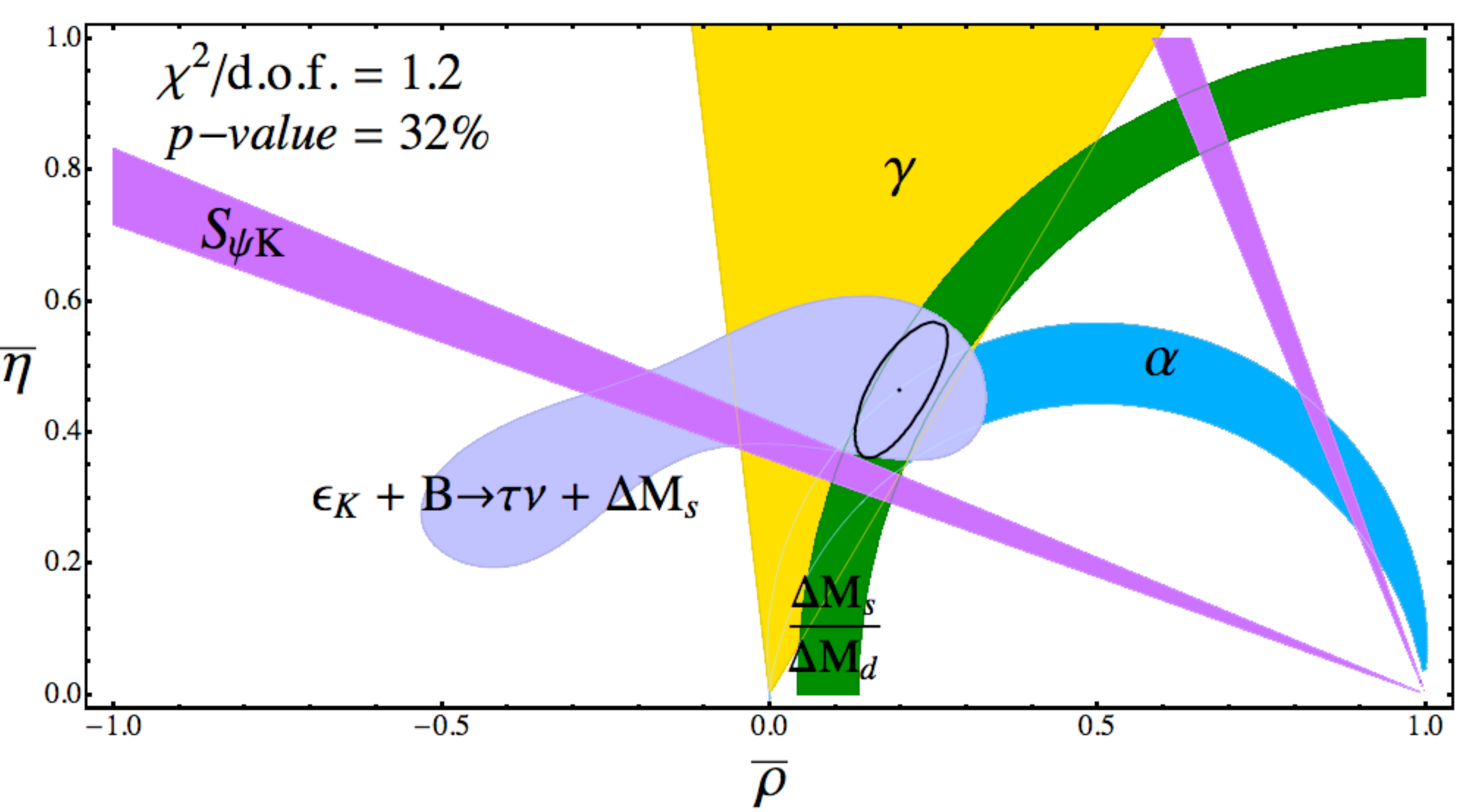}
\caption{{\it Upper panel:} interplay of the $\varepsilon_K$, $\Delta M_{B_s}$ and ${\rm BR} (B\to \tau \nu)$ constraints. {\it Lower panel:} Unitarity triangle fit without semileptonic decays. The contour is obtained using $\varepsilon_K$, $B\to\tau\nu$, $\gamma$, $\Delta M_{B_s}$ and $\Delta M_{B_d}$.\label{fig:utfit}}
\end{center}
\end{figure}
\paragraph{\bf Removing semileptonic decays}
\label{sec:vcb}
Recall that inclusive and exclusive $b\to (c,u) \ell\nu$ decays are tree--level Standard Model processes and, therefore most likely, are quite insensitive to the presence of new physics. The $|V_{cb}|$ constraint translates into a determination of the parameter $A$ of the CKM matrix. Knowledge of the latter is critical in order to extract information from $|V_{ub}| \propto A$, ${\rm BR} (B\to \tau \nu) \propto A^2$ and $\varepsilon_K \propto A^4$ (see Eqs.~(\ref{ek},\ref{btn})). The $(\rho,\eta)$ regions allowed by each of these three observables is obtained {\em with} the inclusion of $|V_{cb}|$. Without any information on $A$ these bands would cover the whole $(\rho, \eta)$ plane. The main role of the determination of $|V_{ub}|$ is to limit the amount of new physics contributions to the phase of $B_d$ mixing; in fact, an upper limit on $|V_{ub}|$ implies an upper limit on $S_{\psi K} = \sin 2 (\beta+\theta_d)$~\cite{Altmannshofer:2009ne,Buras:2009if}. Presently inclusive and exclusive determinations of $|V_{cb}|$ and $|V_{ub}|$ differ at the $2\sigma$ level (see Table~\ref{tab:utinputs}). The exclusion of the $|V_{ub}|$ constraint is not critical any longer to the presence of the $2\sigma$ effects in Eqs.~(\ref{cepsilonfull},\ref{thetadfull}) as emphasized recently in~\cite{Lunghi:2008aa}. In particular the prediction that we obtain for the $B_d$ mixing phase in the no--$V_{ub}$ scenario reads $\left[\sin 2 \beta\right]_{\rm fit} = 0.840 \pm 0.056$ deviating by 2.8$\sigma$ from its direct determination.
%%%UPDATE sigmas%%%
On the other hand, $|V_{cb}|$ appears to be central: employing only its exclusive (inclusive) determination, the $2.0\sigma$ significance of the extraction of $C_\varepsilon$ shifts to $2.5\sigma$ ($1.6\sigma$); similarly the 2.2$\sigma$ effect in $B_d$ mixing shifts to $2.9\sigma$ ($1.4\sigma$).

We now come to elaborating on the new approach that we are advocating here in which no use of semi-leptonic decays will be made. Note that the critical issue is the determination of $A$ from $|V_{cb}|$. We find that the interplay of $\varepsilon_K$, ${\rm BR} (B\to \tau\nu)$ and $\Delta M_{B_s}$ results in a fairly strong constraint on the $(\rho,\eta)$ plane even without using semileptonic decays at all. A simple way to understand this result is to use Eqs.~(\ref{dmbs}-\ref{btn}) to eliminate $A$ and write:
\bea
|\varepsilon_K| &\propto & \hat B_K \;( f_{B_s} \hat B_s^{1/2})^{-4} \label{ekdmbs}\\
|\varepsilon_K| &\propto & \hat B_K \;  {\rm BR} (B\to \tau \nu)^2 \; f_B^{-4} \label{ekbtn}
\eea
where for simplicity we kept only the dominant contributions to $\varepsilon_K$ (proportional to $A^4$) and did not explicitly write the dependence of $\varepsilon_K$ on $\rho$, $\eta$ and all other quantities that are irrelevant to the error budget. Eqs.~(\ref{ekdmbs}) and (\ref{ekbtn}) show that the $|V_{cb}|$ constraint can be effectively replaced by either $f_{B_s} \hat B_s^{1/2}$ or ${\rm BR} (B\to \tau\nu) \times f_B^{-2}$. 
%%%UPDATE p-values, sin2beta, sigma, vcb %%%
In the upper and lower panels of Fig.~\ref{fig:utfit} we show respectively the anatomy of the $\varepsilon_K$, $\Delta M_{B_s}$ and ${\rm BR} (B\to \tau \nu)$ constraints and the complete fit of the unitarity triangle in absence of semileptonic decays. The fit results for $|V_{ub}|$ and ${\rm BR} (B\to \tau\nu)$ do not deviate significantly from Eqs.~(\ref{vubfit}-\ref{brbtnfit}); the extracted value of $\left[\sin 2 \beta\right]_{\rm fit} = 0.811 \pm 0.074$ deviates by 1.8$\sigma$ from its direct determination. It is interesting to observe that the result $|V_{cb}|_{\rm fit}  =  (43.2 \pm 0.9) \times 10^{-3}$ is slightly larger than the average we quote in Table~\ref{tab:utinputs}: this is yet another manifestation of the tension between $K$ and $B_d$ mixing that we are observing in the fit. Finally, we note that $f_{B_s} \hat B_s^{1/2}$ and $\xi = f_{B_s} \hat B_s^{1/2}/f_{B_d} \hat B_d^{1/2}$ are largely independent because they are affected by different lattice systematics and we average results from different lattice collaborations thereby reducing the possible correlation between statistical errors. A surprising outcome is the slight preference of the fit for new physics in $B_d$ mixing. This can be seen by extracting $C_\varepsilon$, $\theta_d$ and $r_H$
\begin{figure}[t]
\begin{center}
\includegraphics[width=0.85 \linewidth]{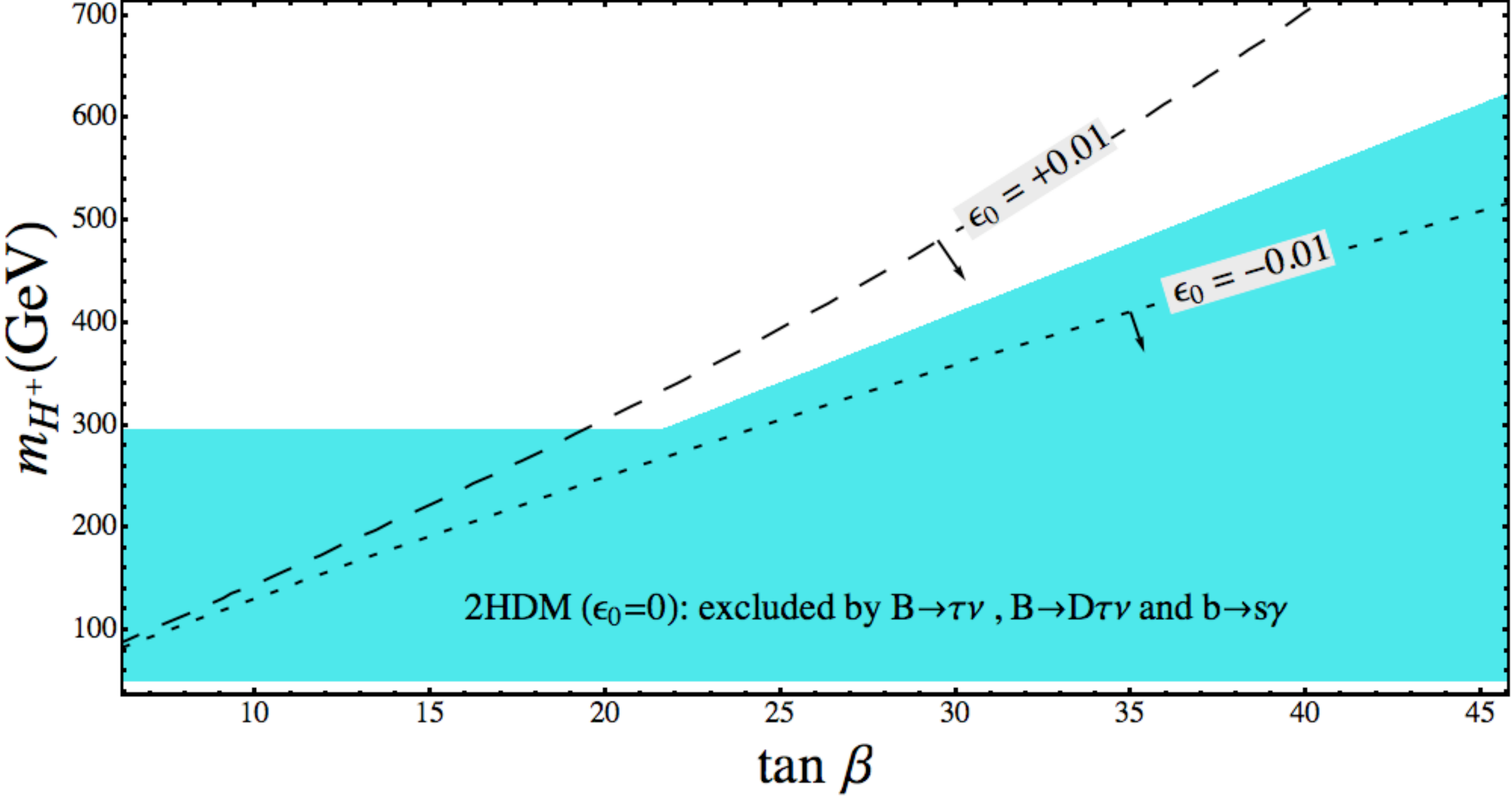}
\caption{95\%C.L. bounds in the $(\tan\beta,m_{H^+})$ plane. The shaded region is excluded in the 2HDM. The dotted (dashed) line shows how this region shifts in two MSSM scenarios with $\epsilon_0 = -0.01\; (0.01)$ (the arrows indicate the excluded region). \label{fig:tanbeta-mhp}}
\end{center}
\end{figure}
%
%
%%%UPDATE%%%
\bea
C_\varepsilon^{{\rm no}V_{qb}} & = & 1.23 \pm 0.30 \hskip 0.35 cm \Rightarrow \; (0.8\sigma, p = 39\%)  , \\
\theta_d^{{\rm no}V_{qb}} & = & -(8.4 \pm 4.6)^{\rm o} \Rightarrow   \; (1.8\sigma, p = 87\%) , \\
r_H^{{\rm no}V_{qb}} & = & 1.7 \pm 0.5 \hskip 0.72cm \Rightarrow \; (1.4\sigma, p =64\%)  . \label{rhpred}
\eea
and noting that new physics in $B_d$ mixing yields larger p--value than new physics in $K$ mixing or in $B\to \tau\nu$.

As an illustration of the implications of these constraints we consider the impact on two Higgs doublet models. Within these models the $r_H \neq 1 $ result can be translated into a constraint on the mass of charged Higgs. In the type-II two Higgs Doublet Model (2HDM) and in the Minimal Supersymmetric SM (MSSM) we can write~\cite{Hou:1992sy} $r_H =  (1 - X_H )^2$ where $X_H = (\tan\beta m_{B^+} / m_{H^+})^2 /(1 + \epsilon_0 \tan\beta)$, $\tan\beta$ is the ratio of the vacuum expectation values of the Higgses that couple to up and down quarks and $\epsilon_0$ summarizes supersymmetric corrections to the $\bar ubW^+$ vertex. In the 2HDM we have $\epsilon_0 = 0$; in the MSSM $\epsilon_0$ does not vanish and typical values range at the $10^{-2}$ level. A full supersymmetric analysis of Eq.~(\ref{rhpred}) is beyond the scope of this letter. In Fig.~\ref{fig:tanbeta-mhp} we present the regions of the $(\tan\beta,m_{H^+})$ that are allowed at 95\% C.L. for various values of $\epsilon_0$. In addition to the bounds implied by Eq.~(\ref{rhpred}) we include also constraints from $B\to D\tau\nu$~\cite{Nierste:2008qe} and $B\to X_s \gamma$~\cite{Misiak:2006zs}. From the observation that $X_H$ is always positive follows that the charged Higgs exchange can only reduce the $B\to\tau\nu$ branching ratio unless $X_H > 2$ implying a sign switch in the $B\to \tau \nu$ amplitude. Eq.~(\ref{rhpred}) implies $X_H = (2.3 \pm 0.2) \vee (-0.3 \pm 0.2)$. At the 1$\sigma$ level only the $X_H \sim 2$ solution is permitted (remember that $X_H > 0$) and the resulting allowed narrow band at low $M_{H^+}$ or large $\tan\beta$, is, in turn, excluded by $B\to D\tau\nu$ data both in the 2HDM and in the MSSM (we follow the numerical analysis of Ref.~\cite{Nierste:2008qe}). At 95\% C.L. the solution $X_H =0$ opens up, corresponding to large $M_{H^+}$. In the 2HDM the $B\to X_s \gamma$ constraint implies $m_{H^+} > 295~{\rm GeV}$~\cite{Misiak:2006zs}. In the MSSM, chargino loops contributions to the $b\to s\gamma$ amplitude can compensate charged Higgs effects: the bound on the charged Higgs depends strongly on the chosen point in the supersymmetric parameter space~\cite{Barenboim:2007sk}. 

Let us now discuss the dominant sources of uncertainties in this analysis. In the following table we list the most relevant inputs, their errors and their impact on $\varepsilon_K$ (as it follows from Eqs.~(\ref{ek}), (\ref{ekdmbs}) and (\ref{ekbtn})):
%
%%%UPDATE if error on Bs mixing from lattice changes %%%
\begin{center}
\begin{tabular}{|c||c||c||c||c|c||}\hline
$X:$ & $\hat B_K$ & $|V_{cb}|$& $f_{B_s} \hat B_s^{1/2}$ & ${\rm BR} (B\to \tau\nu)$ & $f_B$ \cr\hline
$\delta X:$ & $4\%$& $2.5\%$ & $6.9\%$ & $26\%$ & $5\%$   \cr 
$\delta \varepsilon_K:$ & $4\%$& $10\%$  & $27.6\%$ & $52\%$ & $20\%$  \cr\hline
\end{tabular}
\end{center}
\begin{figure}[t]
\begin{center}
\includegraphics[width=0.85 \linewidth]{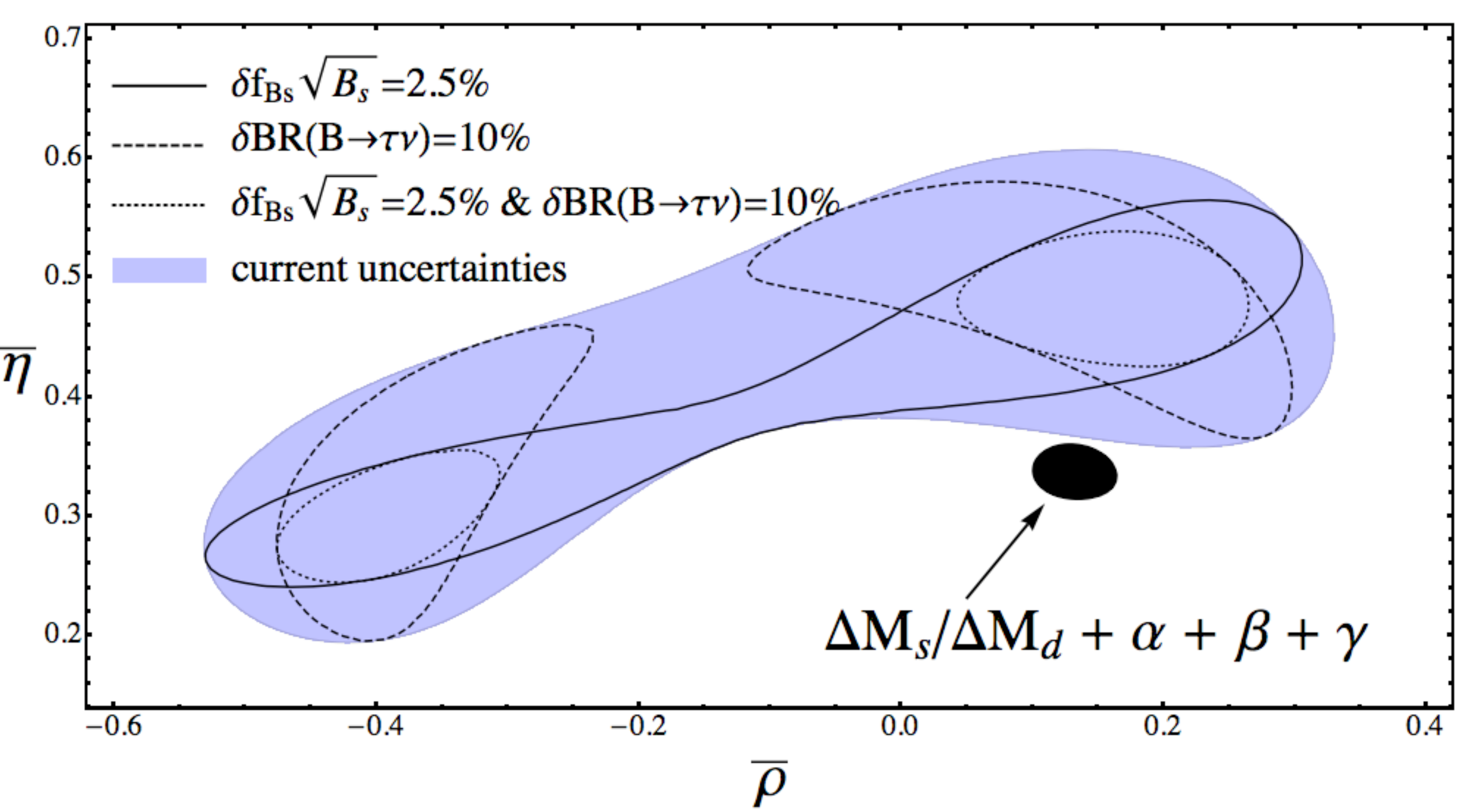}
\caption{Reducing the errors on $f_{B_s} \hat B_s^{1/2}$ and ${\rm BR} (B\to\tau\nu)$. \label{fig:utfit-error}}
\end{center}
\end{figure}
First of all, note that the impact of $\hat B_K$ on the error is subdominant. The use of the semileptonic $b\to c$ constraint results in a $\sim 10\%$ determination of $\varepsilon_K$, roughly half of the uncertainty obtained by employing only $\Delta M_{B_s}$ (i.e.: $f_{B_s} \hat B_s^{1/2}$). A calculation of $f_{B_s} \hat B_s^{1/2}$ at the 2.5\% level would reduce the overall uncertainty on $\varepsilon_K$ to $10\%$; a calculation at the $1\%$ level would impact $\varepsilon_K$ at the same level as $\hat B_K$. At first sight, the impact of $B\to \tau\nu$ seems irrelevant. Fortunately the non--trivial dependence of ${\rm BR} (B\to \tau\nu)$ on $\rho$ and $\eta$ implies a certain degree of orthogonality between the constraints (\ref{ekdmbs}) and (\ref{ekbtn}), as can be seen explicitly in the upper panel of Fig.~\ref{fig:utfit}. A numerical estimate of the impact of this constraint can be obtained by removing it from the fit and recalculating the overall p--value: we obtain $p = 43\%$, meaning that no hint of new physics is observed. The experimental uncertainty on the $B\to \tau \nu$ branching ratio is therefore an important ingredient of this analysis. Once the latter reaches the $10\%$ level, improvements on $f_B$ will be relevant as well. We summarize this discussion in Fig.~\ref{fig:utfit-error} and in the following table:
%
%%%UPDATE%%%
\begin{center}
\begin{tabular}{|cc||c||ccc|}\hline
$\delta_\tau$ & $\delta_s$ & $p_{\rm SM}$ & $\theta_d \pm \delta\theta_d$&$p_{\theta_d}$  & $\theta_d/\delta\theta_d$\cr \hline\hline
${}^{*}26\%$ & ${}^{*}6.8\%$ & $32\%$ & $-(8.4 \pm 4.6)^{\rm o}$ & $87\%$&$1.8\sigma$ \cr\hline \hline
${}^{*}26\%$ & $2.5\%$ & $3.3\%$ & $-(9.6 \pm 3.5)^{\rm o}$ & $85\%$&$2.7\sigma$\cr
${}^{*}26\%$ & $1\%$ & $0.1\%$ & $-(10.0 \pm 3.0)^{\rm o}$ & $84\%$&$3.4\sigma$\cr\hline\hline
$10\%$ & ${}^{*}6.8\%$ & $2\%$ & $-(8.7 \pm 2.8)^{\rm o}$ & $87\%$&$3.1\sigma$\cr
$3\%$ & ${}^{*}6.8\%$ & $0.08\%$ & $-(8.7 \pm 2.3)^{\rm o}$ & $87\%$&$3.8\sigma$\cr\hline\hline
$10\%$ & $2.5\%$ & $0.1\%$ & $-(9.2 \pm 2.5)^{\rm o}$ & $84\%$&$3.7\sigma$\cr
$10\%$ & $1\%$ & $0.004\%$ &  $-(9.6 \pm 2.2)^{\rm o}$ & $83\%$&$4.3\sigma$\cr
$3\%$ & $2.5\%$ & $0.004\%$ &  $-(9.1 \pm 2.1)^{\rm o}$ & $84\%$&$4.4\sigma$\cr 
$3\%$ & $1\%$ & $0.00009\%$ &  $-(9.4 \pm 1.9)^{\rm o}$ & $82\%$&$5.0\sigma$\cr\hline
\end{tabular}
\end{center}
where $\delta_\tau = \delta {\rm BR} (B\to\tau\nu)$, $\delta_s = \delta (f_{B_s} \hat B_s^{1/2})$ and ${}^{*}$ denotes the current uncertainties. The values $\delta_\tau = (10,3)\%$ correspond to a super--$B$ factory result with $(5,50){\rm ab}^{-1}$~\cite{hints09}. In the table we show the p--value of the SM fit and the result of the analysis of the scenario with new physics in $B_d$ mixing (we display the new physics phase $\theta_d$, the p--value of the new physics fit and its significance). We do not show the scenarios with new physics in $K$ mixing or $B\to\tau\nu$ because they yield very low confidence levels and are, therefore, disfavored. From the inspection of the table, we conclude that even modest improvements in $f_{B_s} \hat B_s^{1/2}$ and/or ${\rm BR} (B\to\tau\nu)$ will help enormously in isolating the presence of new physics in the unitarity triangle fit. 
%%%UPDATE p-value and sigma %%%%
For comparison, reducing the total uncertainty on $|V_{cb}|$ to the 1\% level, yields $p_{\rm SM} =  2.8\%$ corresponding to 2.7$\sigma$ effects in either $\theta_d$ or $C_\epsilon$.
\paragraph{\bf Conclusions}
\label{sec:conclusions}
The tension in the standard fit of the unitarity triangle can be interpreted as a hint for new physics. The $\sim 2\sigma$ discrepancies in the extraction of $|V_{cb}|$ and $|V_{ub}|$ between inclusive and exclusive semileptonic $B$ decays tend to cast doubt on the reliability of this conclusion. We investigated the removal of these constraints from the fit. In contrast with the generally accepted statement that information on $|V_{cb}|$ is required in order to make meaningful use of $\varepsilon_K$, we showed, for the first time that, the combination of $\varepsilon_K$, $\Delta M_{B_s}$ and ${\rm BR} (B\to \tau \nu)$ provide quite a stringent constraint in the $(\rho,\eta)$ plane. 

%%%UPDATE sigmas %%%
After removing information from semileptonic decays, we find that the tension in the unitarity triangle fit survives and can be translated into a $1.8\sigma$ hint for new physics in $B_d$ mixing. The preference of the fit for new physics contributions to $B_d$ mixing is caused by the $B\to\tau\nu$ constraint. This branching ratio is proportional to $|V_{ub}|^2$ and, as it can be seen from Fig.~\ref{fig:utfit}, points to a large value of $|V_{ub}/V_{cb}|$. This, in turn, favors a scenario with new physics in the $B_d$ mixing phase. 

Even modest improvements on ${\rm BR} (B\to \tau\nu)$ and/or $f_{B_s} \hat B_s^{1/2}$ may push this tension above the $3\sigma$ level; if errors on both constraints are reduced simultaneously, $\delta_\tau = (10,3)\%$ and $\delta_s = (2.5,1)\%$, the effect reaches $(4-5)\sigma$. Note that improvements on $B\to\tau\nu$ require a super--$B$ factory~\cite{Akeroyd:2004mj,Bona:2007qt,Browder:2008em} while the reduction of $\delta_s$ is a purely theoretical ({\it i.e.} lattice) issue.  Note also that lattice calculations of matrix elements relevant for decay constants or for ($K^0$, $B_d$, $B_s$) oscillations and therefore for $\delta_s$, do not require momentum injection unlike the calculation of the semileptonic form factors and to that extent are simpler. However, we stress again that we are not suggesting abandoning the traditional approach with use of semileptonic decays, but rather in addition making concerted efforts towards improved lattice determination of $f_{B_s} \hat B_s^{1/2}$ and also of the ${\rm BR}(B\to \tau \nu)$. These should provide valuable redundancy in our quest for new physics through flavor studies even in the LHC era. Finally we would like to stress that the main focus of the present letter is to propose a new clean strategy to implement simultaneously $K$ and $B_d$ mixing constraints on the $(\rho,\eta)$--plane and that our projections on the reach of this method depend solely on improving $\delta_\tau$ and $\delta_s$ and are quite insensitive to the rest of the inputs summarized in Table~\ref{tab:utinputs}, in particular, the assumed errors in lattice computations.
\begin{acknowledgments}
This research was supported in part by the U.S. DOE contract No.DE-AC02-98CH10886(BNL).
\end{acknowledgments}


\begin{thebibliography}{99}
\bibitem{footnote1}
In this letter, for clarity, only constraints coming from $b\to (c,u) \ell \nu$ with $\ell = e,\mu$ are called direct determinations of $|V_{cb}|$ and $|V_{ub}|$.
\bibitem{footnote2}
We remind the reader that the p--value is defined as the probability to obtain a minimum chi-square larger than the observed one, assuming the SM is correct (e.g. $p=15\%$ means that there is a 15\% chance that the observed result is explained by fluctuations).
\end{thebibliography}
\end{document}